# HANDBOOK FOR THE GREAT08 CHALLENGE: AN IMAGE ANALYSIS COMPETITION FOR COSMOLOGICAL LENSING

By Sarah Bridle,[1,20] John Shawe-Taylor,[1] Adam Amara,[2] Douglas Applegate,[3] Sreekumar T. Balan,[1] Joel Berge,[4,5,6] Gary Bernstein,[7] Hakon Dahle,[8] Thomas Erben,[9] Mandeep Gill,[10] Alan Heavens,[11] Catherine Heymans,[12,19,21] F. William High,[13] Henk Hoekstra,[14] Mike Jarvis,[7] Donnacha Kirk,[1] Thomas Kitching,[15] Jean-Paul Kneib,[8] Konrad Kuijken,[16] David Lagatutta,[17] Rachel Mandelbaum,[18] Richard Massey,[5] Yannick Mellier,[19] Baback Moghaddam,[4,5] Yassir Moudden,[6] Reiko Nakajima,[7] Stephane Paulin-Henriksson,[6] Sandrine Pires,[6] Anais Rassat,[6] Alexandre Refregier,[6] Jason Rhodes,[4,5,22] Tim Schrabback,[16] Elisabetta Semboloni,[9] Marina Shmakova,[3] Ludovic van Waerbeke,[12] Dugan Witherick,[1] Lisa Voigt[1] and David Wittman[17]

*[1]University College London, [2]University of Hong Kong, [3]Stanford Linear Accelerator Center, [4]Jet Propulsion Laboratory, [5]California Institute of Technology, [6]Commissariat a l'Energie Atomique, Saclay, [7]University of Pennsylvania, [8]Laboratoire d'Astrophysique de Marseille, [9]University of Bonn, [10]Ohio State University, [11]Royal Observatory, University of Edinburgh, [12]University of British Columbia, [13]Harvard University, [14]University of Victoria, [15]University of Oxford, [16]University of Leiden, [17]University of California, Davis, [18]Institute for Advanced Study, Princeton and [19]Institut d'Astrophysique de Paris*

The GRavitational lEnsing Accuracy Testing 2008 (GREAT08) Challenge focuses on a problem that is of crucial importance for future observations in cosmology. The shapes of distant galaxies can

---

Received April 2008; revised October 2008.
[20]Supported by the Royal Society in the form of a University Research Fellowship.
[21]Supported by a European Commission Programme 6th framework Marie Curie Outgoing International Fellowship under Contract MOIF-CT-2006-21891.
[22]Supported in part by the Jet Propulsion Laboratory, which is run by Caltech under a contract from NASA.
*Key words and phrases.* Inference, inverse problems, astronomy.

be used to determine the properties of dark energy and the nature of gravity, because light from those galaxies is bent by gravity from the intervening dark matter. The observed galaxy images appear distorted, although only slightly, and their shapes must be precisely disentangled from the effects of pixelisation, convolution and noise. The worldwide gravitational lensing community has made significant progress in techniques to measure these distortions via the Shear TEsting Program (STEP). Via STEP, we have run challenges within our own community, and come to recognise that this particular image analysis problem is ideally matched to experts in statistical inference, inverse problems and computational learning. Thus, in order to continue the progress seen in recent years, we are seeking an infusion of new ideas from these communities. This document details the GREAT08 Challenge for potential participants. Please visit www.great08challenge.info for the latest information.

**1. Introduction.** Our Universe appears to be dominated by dark matter and dark energy [Biello and Caldwell (2006), Linder and Perlmutter (2007)]. These are not well described or even understood by modern science, so studying their properties could provide the next major breakthrough in physics. This may ultimately lead to a discovery of a new class of fundamental particle or a theory of gravity that supersedes Einstein's theory of general relativity. For this reason, the primary science drivers of most cosmological surveys are the study of dark matter and dark energy. Funding agencies worldwide have committed substantial resources to tackling this problem; several of the planned projects will spend tens to hundreds of millions of taxpayers' Euros on this topic.

Many cosmologists have concluded that gravitational lensing holds the most promise to understand the nature of dark matter and dark energy [Albrecht et al. (2006), Peacock et al. (2006)]. Gravitational lensing is the process in which light from distant galaxies is bent by the gravity of intervening mass in the Universe as it travels towards us. This bending causes the shapes of galaxies to appear distorted [Bartelmann and Schneider (2001), Wittman (2002), Refregier (2003b) and Munshi et al. (2006)]. We can relate measurements of the statistical properties of this distortion to those of the dark matter distribution at different times in the history of the Universe. From the evolution of the dark matter distribution we can infer the main properties of dark energy.

To extract significant results for cosmology, it is necessary to measure the distortion to extremely high accuracy for millions of galaxies, in the presence of observational problems such as blurring, pixelisation and noise and theoretical uncertainty about the undistorted shapes of galaxies. Our techniques are good enough to analyse current data but we need a factor of ten improvement to capitalise on future surveys, which requires an injection of new ideas and expertise. We challenge you to solve this problem.

Section 2 explains the general problem and presents an overview of our current methods. Section 3 describes in detail the GREAT08 Challenge simulations, rules and assessment. We conclude in Section 4 with a summary of the additional issues that arise in more realistic image analysis, that could be the basis of future GREAT Challenges.

**2. The problem.** For the vast majority of galaxies the effect of gravitational lensing is to simply apply a matrix distortion to the whole galaxy image

$$\begin{pmatrix} x_{\mathrm{u}} \\ y_{\mathrm{u}} \end{pmatrix} = \begin{pmatrix} 1-g_1 & -g_2 \\ -g_2 & 1+g_1 \end{pmatrix} \begin{pmatrix} x_{\mathrm{l}} \\ y_{\mathrm{l}} \end{pmatrix}, \tag{2.1}$$

where a positive "shear" $g_1$ stretches an image along the $x$ axis and compresses along the $y$ axis; a positive shear $g_2$ stretches an image along the diagonal $y = x$ and compresses along $y = -x$. The coordinate $(x_{\mathrm{u}}\ y_{\mathrm{u}})$ denotes a point on the original galaxy image (in the absence of lensing) and $(x_{\mathrm{l}}\ y_{\mathrm{l}})$ denotes the new position of this point on the distorted (lensed) image. There is also an isotropic scaling that we ignore here. This seems a sensible parameterisation to use for the shear because it is linear in the mass [e.g., Kaiser, Squires and Broadhurst (1995)]. The top left two panels of Figure 2 illustrate an exceptionally high quality galaxy image before and after application of a large shear. For cosmic gravitational lensing a typical shear distortion is $g_i \sim 0.03$, therefore a circular galaxy would appear to be an ellipse with major to minor axis ratio of 1.06 after shearing. Note that the three-dimensional shape of the galaxy is not important here; we are concerned only with the two-dimensional (projected) shape.

Since most galaxies are not circular, we cannot tell whether any individual galaxy image has been sheared by gravitational lensing. We must statistically combine the measured shapes of many galaxies, taking into account the (poorly known) intrinsic galaxy shape distribution, to extract information on dark matter and dark energy. Shear correlations were first measured in 2000 [Bacon, Refreiger and Ellis (2000), Kaiser, Wilson and Luppino (2000), Wittman et al. (2000) and van Waerbeke et al. (2000)] and the most recent results [Massey et al. (2007c), Fu et al. (2008)] use millions of galaxies to measure the clumpiness of dark matter to around 5 percent accuracy. Figure 1 shows a three-dimensional map of the dark matter reconstructed by Massey et al. (2007b). Future surveys plan to use roughly a billion galaxies to measure the dark matter clumpiness to extremely high accuracy and thus measure the properties of dark energy to 1 percent accuracy. This will require a measurement accuracy on each of $g_1$ and $g_2$ of better than 0.0003. However this can only be achieved if statistical inference problems can be overcome.

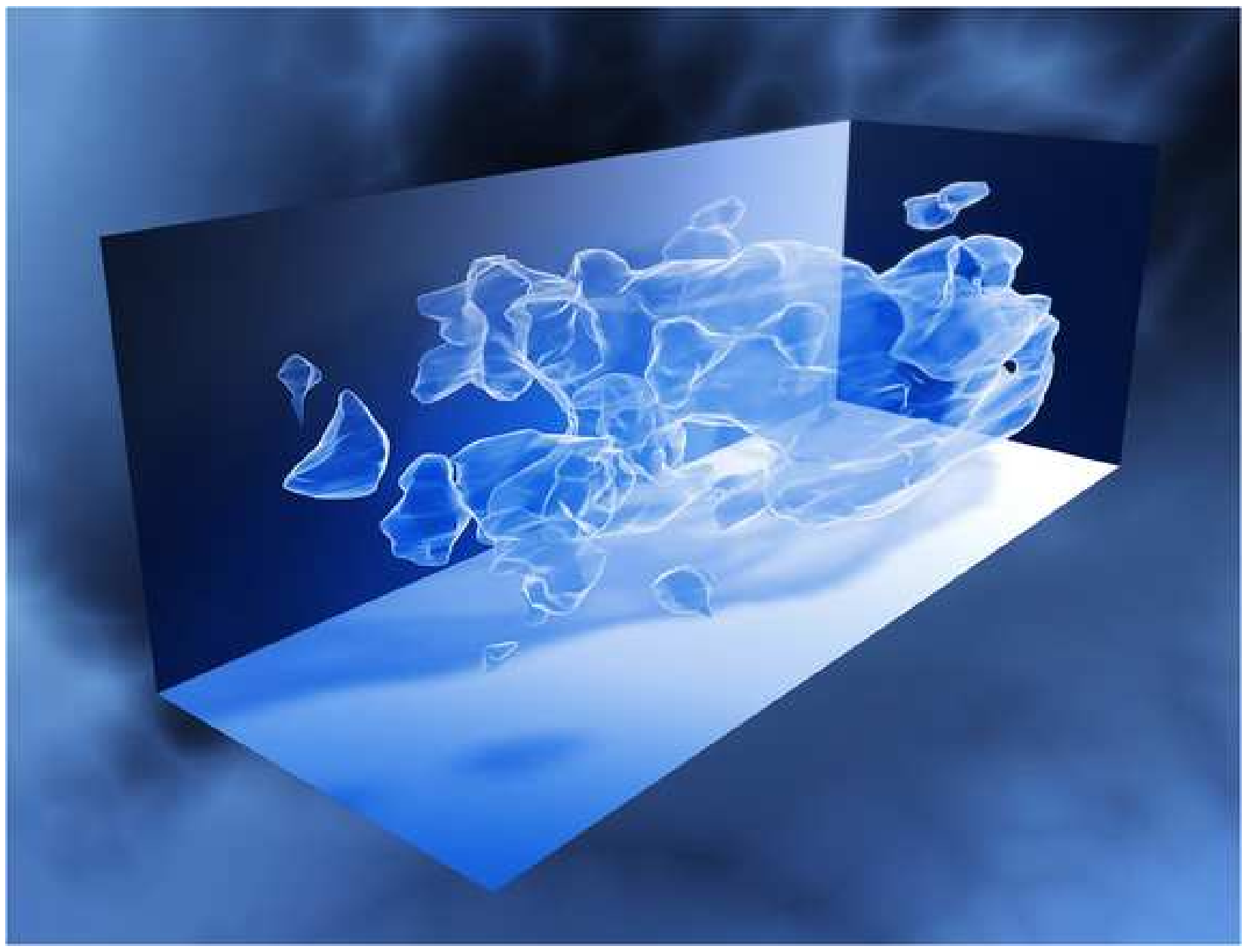

Fig. 1. *Illustration of the invisible dark matter distribution inferred using gravitational lensing detected in the Cosmic Evolution Survey (COSMOS) [Massey et al. (2007b)]. The three axes of the box correspond to sky position (in right ascension and declination), and distance from the Earth increasing from left to right (as measured by cosmological redshift). Image credit: NASA, ESA and R. Massey (California Institute of Technology).*

Shear measurement is an inverse problem, illustrated in Figures 2 and 3. The forward process is illustrated in Figure 2: (i) each galaxy image begins as a compact shape, which appears sheared by the operation in equation (2.1); (ii) the light passes through the atmosphere (unless the telescope is in space) and telescope optics, causing the image to be convolved with a kernel; (iii) emission from the sky and detector noise cause a roughly constant "background" level to be added to the whole image; (iv) the detectors sum the light falling in each square detector element (pixel); and (v) the image is noisy due to a combination of Poisson noise[23] in the number of photons arriving in each pixel, plus Gaussian noise due to detector effects. The majority of galaxies we need to use for cosmological measurements are faint: a typical uncertainty in the total amount of galaxy light is 5 percent.

[23] Poisson noise arises because there is a finite number of photons arriving at the detector during the fixed length of time that the shutter is open. The probability of receiving $n$ photons in a pixel is therefore given by $\Pr(n|\lambda) = \lambda^n e^{-\lambda}/n!$ where $\lambda$ is the mean number of photons observed in that pixel during many exposures of the same length of time.

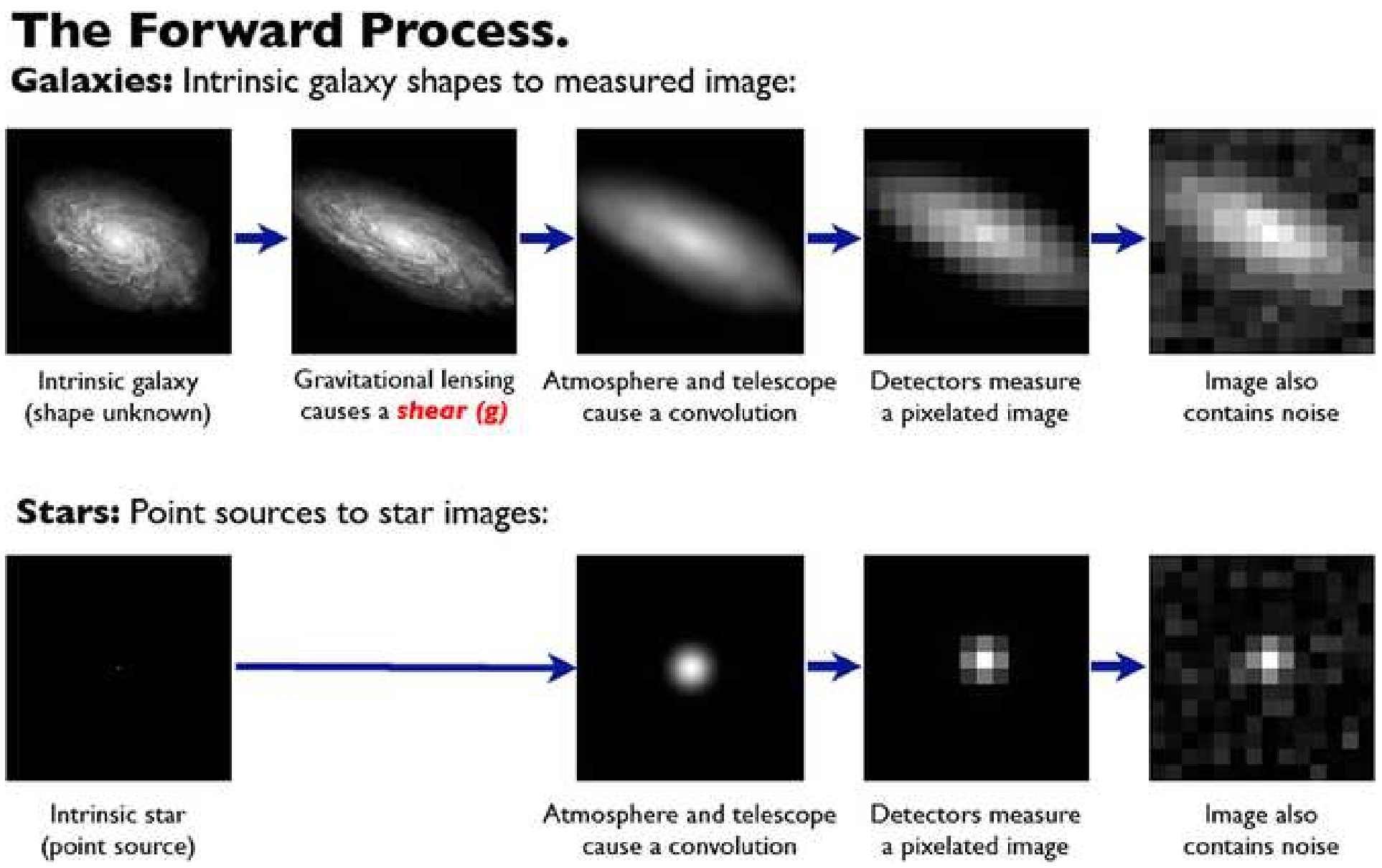

FIG. 2. *Illustration of the forward problem. The upper panels show how the original galaxy image is sheared, blurred, pixelised and made noisy. The lower panels show the equivalent process for (point-like) stars. We only have access to the right hand images.*

Stars are far enough away from us to appear point-like. They therefore provide noisy and pixelised images of the convolution kernel (lower panels of Figure 2). The convolution kernel is typically of a similar size to the galaxies we are observing. If it were not accounted for, we would therefore underestimate the shear. The kernel can also be up to ten times more elliptical than the ellipticity induced by gravitational shear. If this is improperly accounted for, it can masquerade as the cosmological effect we are trying to measure. In real astronomical observations, the kernel varies across a single image containing hundreds of stars and galaxies, and also from one image to the next. Since stars are distributed all over the sky we can use nearby stars on a given image to estimate the kernel for a given galaxy.

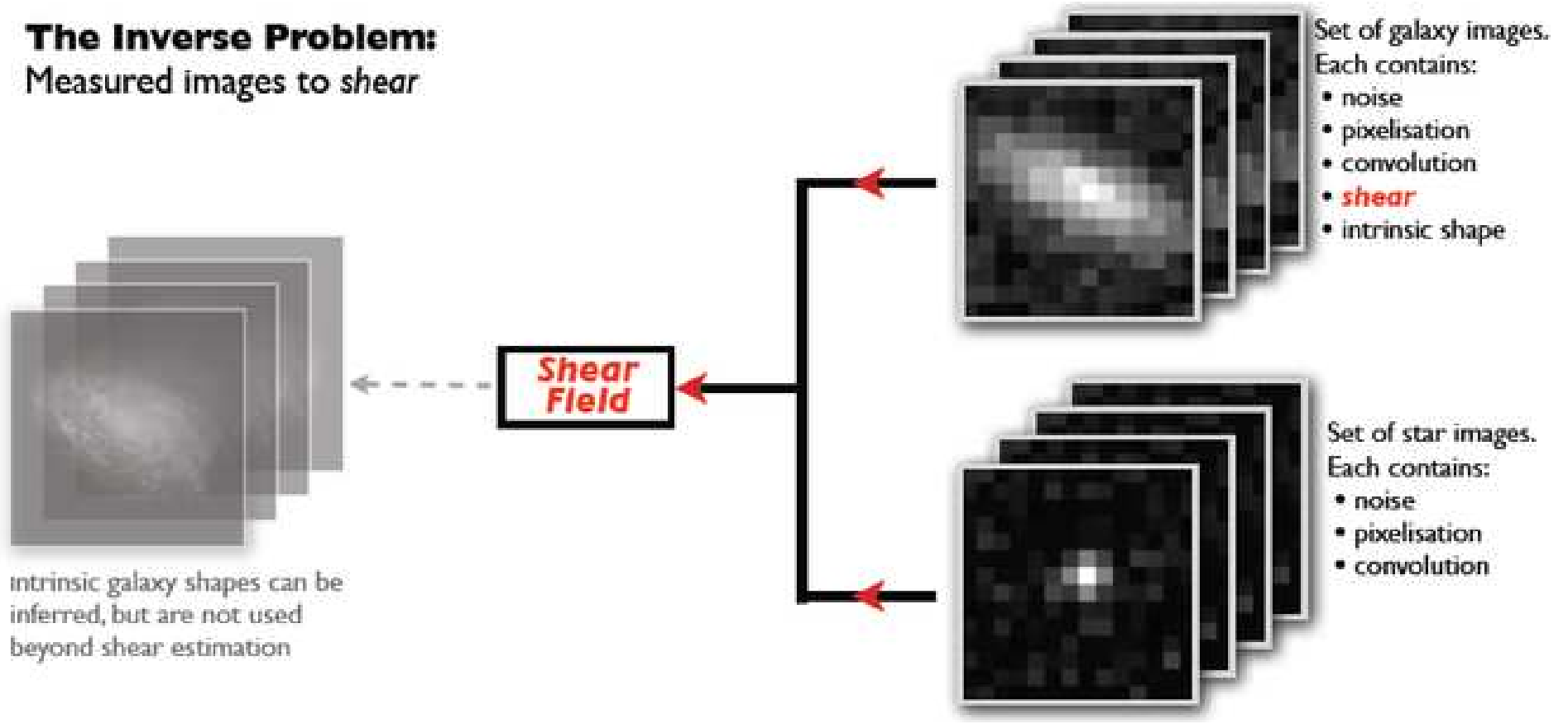

FIG. 3. *Illustration of the inverse problem. We begin on the right with a set of galaxy and star images. The full inverse problem would be to derive both the shears and the intrinsic galaxy shapes. However shear is the quantity of interest for cosmologists.*

The most significant obstacle to shear measurement is that the intrinsic shape of each galaxy is unknown. Even the probability distribution function of possible shapes from which it could have been drawn is highly uncertain; we do not even have a good parameterisation for galaxy shapes. We try to categorise galaxies into three types: spirals (e.g., Figure 2), ellipticals and irregulars but many galaxies are somewhere between the categories.

One good assumption that we can make is that unlensed galaxies are randomly oriented. In addition we find that the radially averaged 1D galaxy light intensity profile $I(r)$ is well fit by $I(r) = I_{\rm o} \exp(-(r/r_{\rm c})^{1/n})$ [Sersic (1968)], where $I_{\rm o}$, $r_{\rm c}$ and $n$ are free parameters and $r$ is the distance from the centre of the galaxy. For elliptical galaxies $n \sim 4$ ("de Vaucouleurs profile") and for spirals $n \sim 1$ ("exponential profile"). Unfortunately we do not have suitable galaxy images which are free of pixelisation and convolution from which to learn about intrinsic galaxy shapes. We can however make low noise observations of some small areas of sky.

Methods developed so far by the lensing community are discussed in detail in the appendices and references therein. At the Challenge launch we will provide code implementing some existing methods. Their performance on earlier blind challenges is discussed in Heymans et al. (2006) and Massey et al. (2007a). In all existing methods each star is analysed to produce some information about the convolution kernel. This is averaged or interpolated over a number of stars to reduce the noise and produce information about the kernel at the position of each galaxy. The galaxy image is analysed, taking into account the kernel, to produce an estimate of the shear ($g_1$ and $g_2$) at the position of that galaxy.

Real astronomical data is simply an image of the continuous night sky. The first step of any analysis pipeline is therefore to identify stars and galaxies (distinguishing small, faint galaxies from small, faint stars in a noisy image is a nontrivial task), cut out images around them and estimate the local background level. Since the convolution kernel also usually varies as a function of time and image position, the apparent shapes of stars must be modelled, and the model coefficients interpolated to the positions of the galaxies. Simplifications have been made in the GREAT08 data to eliminate these steps.

In real data the shear fields $g_1$ and $g_2$ vary across the sky due to the clumpiness of dark matter in the Universe. They also vary with the distance of the galaxy. It is usually reasonable to assume that the shear is constant across the image of a single galaxy. In practice the shear is different for each galaxy but is zero when averaged over a large survey, that is, $\langle g_1 \rangle = \langle g_2 \rangle = 0$. It is necessary to use images of both stars and galaxies to extract the shear field in the presence of the unknown convolution kernel. In this process our priority is *not* to learn about the properties of the unlensed galaxy images.

Conventionally, the shear information from each galaxy image is combined to produce a statistic that can be predicted from a cosmological model. For example, the most common statistic is the shear correlation function $\langle g_{1i} g_{1j} \rangle + \langle g_{2i} g_{2j} \rangle$ [e.g., Bartelmann and Schneider (2001)] where the averages are carried out over all galaxy pairs $i$ and $j$ at a given angular separation on the sky. The properties of dark matter and dark energy can then be inferred by calculating the probability of the observed statistics as a function of cosmological parameters. The whole process is illustrated in Figure 4. Note that GREAT08 focuses entirely on the process of going from image to shear estimate because this is the current bottleneck that is hindering further analysis of astronomical data. However shear measurement methods will ultimately need to fit into this larger scheme to be useful for cosmology.

**3. The GREAT08 Challenge.** In the previous section we described the general cosmic lensing problem. In this section we focus on the specifics of the GREAT08 Challenge. We start by describing the properties of the GREAT08 simulations. We explain how the results are assessed and the winner determined.

3.1. *Simulations.* The Challenge images are made by simulation, using the flowchart of the forward problem (Figure 2). We have made a number of simplifications which we aim to relax for future GREAT Challenges, as discussed in Section 5. The simulations consist of many small (roughly 40 by 40 pixel) images, each containing a single object. The images are clearly labelled as either stars (kernel image) or galaxies. The objects are located roughly, but not exactly, in the centre of each image. The images are divided into different "sets," each containing thousands of images. All the images within a set have identical values of the shear $g_1$ and $g_2$ and an identical convolution kernel. A very large constant is added to all pixels in a set and Poisson noise is added to each pixel. For GREAT08 RealNoise-Known and GREAT08 RealNoise-Blind (see below) the constant is so large that the noise is very close to being Gaussian with the same variance for every pixel in the image. You may use all these facts in your analysis.

The star images in each set provide information on the convolution kernel. To simplify the Challenge we also provide the equations used to make these kernel images. Therefore you have the choice of whether to use the exact equations *or* the star images provided. In each star image the star has a different centre position and different random noise realisation. The noise level and number of star images should be sufficient to reconstruct the convolution kernel to a precision where uncertainties in the convolution kernel are smaller than the small uncertainty due to the finite number of galaxies. Your challenge is to derive an estimate of the shear applied to the galaxy images within each set.

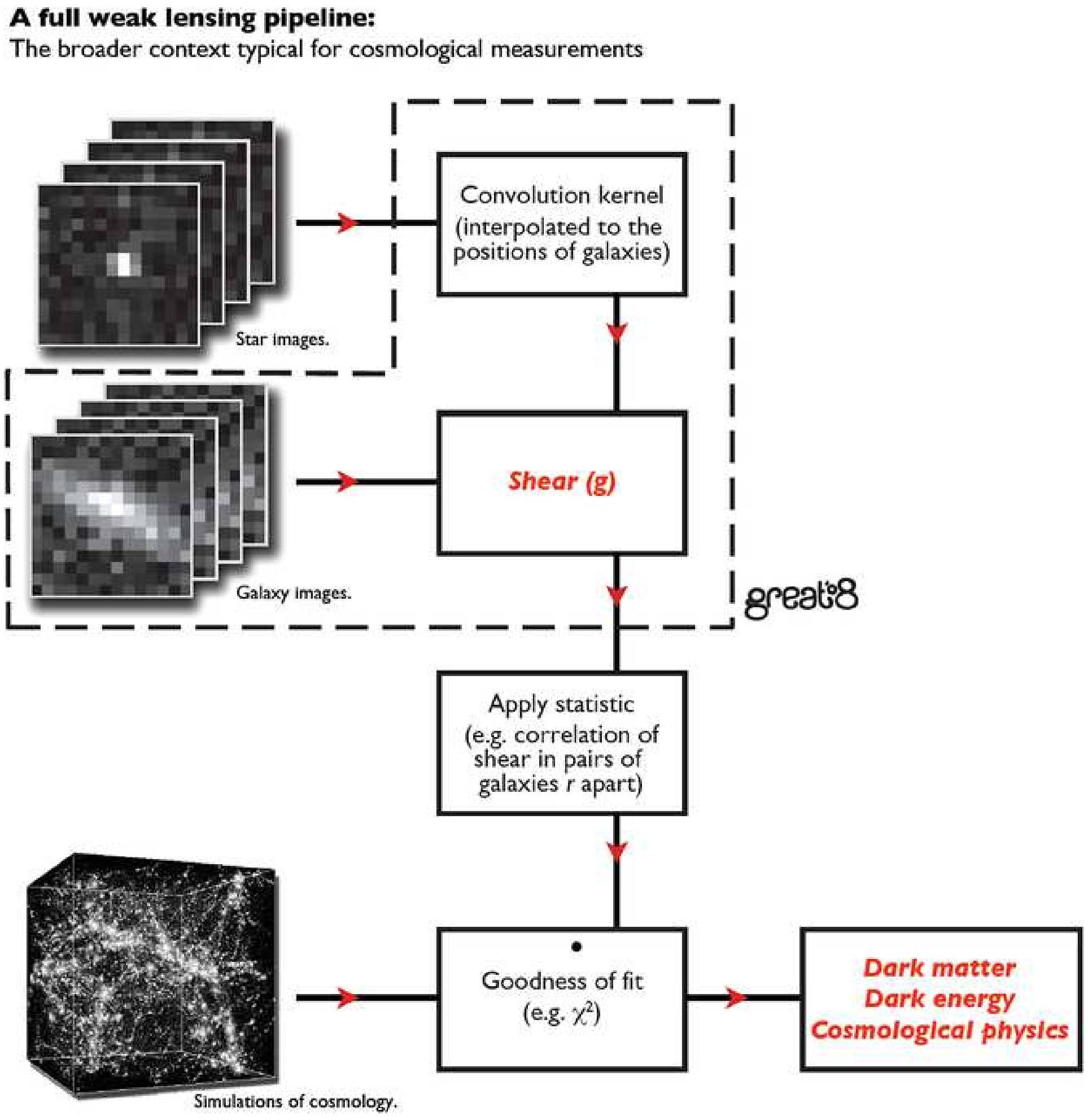

FIG. 4. *Flowchart indicating the extent of a full conventional cosmic gravitational lensing data analysis pipeline, from measuring the convolution kernel using the shapes of stars, to measurements of cosmology. The GREAT08 Challenge focuses exclusively on the steps enclosed in the box made by the dashed black line. The final winners will be determined based solely on estimates of shear. Simulation credit: Andrey Kravtsov (University of Chicago).*

This Challenge is difficult because of the following realistic features: (i) the extremely high accuracy required on the final answer; (ii) a model for the galaxies is not provided, and the galaxy shape and position are different from image to image (drawn from some underlying model which is not disclosed); (iii) there is convolution and noise; (iv) images are pixelised.

The GREAT08 galaxy image simulation types are summarized in Table 1. To make the Challenge more approachable there are a few sets of low noise simulations ("GREAT08 LowNoise-Known" and "GREAT08 LowNoise-Blind"). The true shear values are provided for a subset of these ("GREAT08 LowNoise-Known") and there is a blind competition for the remainder ("GREAT08 LowNoise-Blind"). The main challenge ("GREAT08 RealNoise-Blind") has a realistic, much higher, noise level. There are also some sets with a realistic noise for which the true shears are provided ("GREAT08 RealNoise-Known"). It is not possible to determine the true shear of a galaxy, even with an infinite amount of data. Therefore a method that requires a perfect training set will not be useful in practice. However we will be able to make simulations of the sky using imperfect galaxy models. To simulate this future situation, we use a slightly different model for the galaxies in the "Known" sets than in the "Blind" sets. This means that although methods that require a training set can be used (see rule 4), they may be at a small relatively realistic disadvantage, depending on the sensitivity of the method to the galaxy properties.

3.2. *Results.* Each submission consists of a shear estimate ($g_1$ and $g_2$) for each set of images, with associated 68 percent error bars. A quality factor is calculated for each submission using the differences between the submitted and true shear values.

The goal of the Challenge is to successfully recover the true input shear values used in the simulation, $g^t_{1j}$, $g^t_{2j}$, for each set of images, $j$. You may use whatever method you like to combine the shear information from each galaxy within a set to estimate the shear applied to the whole set. The submitted shear values, $g^m_{1j}$, $g^m_{2j}$, will differ from the true values due to the noise on the images and due to any biases induced by the measurement method. A good method would both filter the noise effectively *and* have small or nonexistent

TABLE 1
*Summary of the three GREAT08 simulation suites*

| | **True shears provided** | **Blind competition** |
|---|---|---|
| Low noise | GREAT08 LowNoise-Known | GREAT08 LowNoise-Blind |
| Realistic noise | GREAT08 RealNoise-Known | GREAT08 RealNoise-Blind |

biases. We define the quality factor in terms of the mean squared error

$$Q = \frac{10^{-4}}{\langle(\langle g^m_{ij} - g^t_{ij}\rangle_{j\in k})^2\rangle_{ik}}, \tag{3.1}$$

where the inner angle brackets denote an average over sets with similar shear value and observing conditions $j \in k$. The outer angle brackets denote an average over simulations with different true shears and observing conditions $k$ and shear components $i$ [see Kitching et al. (2008)].

This deliberately designed to reward methods that have small biases. This is important because in cosmology we average over a very large number of galaxies and any remaining bias will bias cosmological parameters. This definition will also penalise methods that have small biases at the expense of being extremely noisy.

This quantity does not include the error bars you submit. We are not interested in a method which has large but accurate error bars since it will not produce tight cosmological constraints. Furthermore the Challenge images cover only a small (but realistic) range of observing conditions, therefore it is unlikely that an ultimately useful method would lose the competition because of poor performance in a particular corner of observing condition parameter space where your method has particularly large error bars.

It has been shown that a systematic variance $\langle(\langle g^m_{ij} - g^t_{ij}\rangle_{j\in k})^2\rangle_{ik} < 10^{-7}$ will be needed to fully utilise future cosmic lensing data sets [Amara and Refregier (2007)], corresponding to $Q = 1000$ [see also Huterer et al. (2006), Van Waerbeke et al. (2006)]. The number of galaxy images included in GREAT08 LowNoise-Blind and in GREAT08 RealNoise-Blind are sufficient to test $Q$ to this value. If a single constant value of zero shear were submitted ($g^m_{1j} = g^m_{2j} = 0$ for all $j$) then since $\sqrt{\langle g^{t2}_{ij}\rangle_{ij}} \sim 0.03$ it follows that $Q \sim 0.1$. The existing methods that have been used to analyse astronomical data have $Q \sim 10$, which was sufficient for those surveys.

The GREAT08 Challenge Winner is the entry with the highest $Q$ value on GREAT08 RealNoise-Blind. These will be publicly available on a leader board, as mocked-up in Table 2. Results using several existing methods appear on the leader board at the start of the Challenge, to show the current state-of-the art.

The main diagnostic indicator in the leader board is the quality factor $Q$, which determines the ranking of the submissions. As discussed above, the quality factor does not take into account the submitted uncertainty estimates on the shears, whereas an ideal method would calculate these reliably. We make an internal estimate of the uncertainties and compare with your submission to produce an error flag. If the uncertainty estimates are on average wrong to more than a factor of two then this is flagged in the leader board. There are no consequences of the error flag in determining

Table 2
*A mock leader board, showing a potential range of results. Submissions by members of the GREAT08 Team are marked with an asterisk. There are two leader boards: one for GREAT08 LowNoise-Blind and one for GREAT08 RealNoise-Blind*

| **Name** | **Method** | $Q$ | **Error flag** | **Number of submissions** | **Date of last submission** |
|---|---|---|---|---|---|
| A. Einstein | BestLets | 1001 | – | 15 | 25 Dec 2008 |
| Team Bloggs | Joe1 | 582 | Warning | 2 | 2 Nov 2008 |
| Dr. Socrates | ArcheoShapes | 116 | Warning | 212 | 23 Sept 2008 |
| W. Lenser* | KSB+++ | 99 | – | 12 | 10 Aug 2008 |
| A. Monkey | Guess Again | 1.2 | Warning | 5 | 30 Nov 2008 |

the winner. The winner may have an error flag warning and will still win, based on their $Q$ value.

The data for which true shears are provided (GREAT08 LowNoise-Known and GREAT08 RealNoise-Known) are released publically in July 2008. The challenge data (GREAT08 LowNoise-Blind and GREAT08 RealNoise-Blind) are released in fall 2008 and the deadline will be 6 months after the release of the challenge data. Please see www.great08challenge.info for the latest information and discussions in the GREAT08 section of CosmoCoffee at www.cosmocoffee.info. The Challenge deadline is to be followed by a more detailed report making use of the internal structure of the simulations to identify which observational conditions favour which method. We hope this will lead to a publication and workshop.

**4. Conclusions.** The field of cosmic gravitational lensing has recently seen great successes in measuring the distribution of dark matter. Indeed, hundreds of millions of Euros will soon be spent on exciting new surveys to determine the nature of the two fundamental (yet quite mysterious) materials that are the most common in our Universe. Uniquely among cosmological tests, measurements of cosmic lensing are not currently limited by complicated astrophysical processes occurring half-way across the Universe, but by improved techniques for statistical image analysis right here on Earth. Cosmologists have hosted several shear measurement competitions amongst themselves, and developed several methods that achieve an accuracy of a few percent. However, reaching the accuracy required by future surveys needs a fresh approach to the problem. The GREAT08 Challenge is designed to seek out your expertise. Aside from the academic interest in solving a challenging statistical problem, successful methods are absolutely essential for further advances in cosmological investigations of dark matter and dark energy.

GREAT08 marks the first time that the challenge of high precision galaxy shape measurements has been set outside the gravitational lensing community, and as such marks a first step in a global effort to develop the next generation of cosmological tools using expertise, experience and techniques coming from a broad disciplinary base. The field of gravitational lensing is expected to grow at an increasing rate over the coming decade but an injection of new ideas is vital if we are to take full advantage of the potential of lensing to be the most powerful cosmological probe. The GREAT Challenges can therefore be seen as a comprehensive series where the goal of each step is both to bring new insight and to tackle more complicated problems than the previous step.

**5. GREAT08 simplifications and future challenges.** The GREAT08 Challenge outlined in this document is a difficult challenge despite the simplifying assumptions which include:

- Constant Shear: Within each set of images the shear is constant whereas in real data shear is a spatially varying quantity from which correlation statistics are used to measure properties of the Universe.
- Constant Kernel: Within each set the convolution kernel is constant whereas in real data this is a spatially and time varying quantity that also needs to be measured and interpolated between galaxy positions.
- Simple Kernel: The convolution kernels used in this Challenge are simple relative to those of real telescopes.
- Simple Galaxy Shape: The galaxies used in this Challenge are simple relative to real data.
- Simple Noise Model: The noise is Poisson. In practice there would be unusable bad pixels which may be flagged and the noise would be a combination of Gaussian and Poisson, with the Gaussian contribution potentially varying across the image.
- Image Construction: In GREAT08 there is only one object in each small image and each is labelled according to whether it is a star or a galaxy. The selection of galaxies in a real image must not depend on the applied shear otherwise this introduces an additional bias: if very elliptical galaxies are preferentially downweighted then galaxies that happen to be aligned with the shear will tend to be lost which will bias the measured shear low. In addition, in real data some galaxies overlap and are best discarded from conventional analyses. Furthermore, conventional analyses rely on accurate labelling of stars and galaxies.

In GREAT09 we anticipate that many of these assumptions would be relaxed therefore methods which perform well in GREAT08 by overly capitalising on the simplifications may not perform well in GREAT09.

Beyond GREAT09 there are a multitude of further issues that have a significant effect on accurate shape measurement. Cosmic rays and satellite tracks contaminate the image [see Storkey et al. (2004)]; detector pixels vary in responsivity and the responsivity is not linear with the number of photons (Charge Transfer Efficiency); the detector elements are not perfectly square and/or are not perfectly aligned in the telescope so that the sky coordinates do not perfectly map to pixel coordinates, and they bleed (Inter Pixel Responsivity); there are multiple exposures of each patch of sky, each with a different kernel.

The ultimate test and verification of a method will be in its application to data. The goal of the GREAT Challenges is to encourage the development of methods which will one day be used in conjunction with state-of-the-art data in order to answer some of our most profound and fundamental questions about the Universe.

## APPENDIX A: RULES

1. Participants may use a pseudonym or team name on the results leader board, however real names (as used in publications) must be provided where requested during the result submission process.
2. Participants who have investigated several algorithms may enter once per method. Changes in algorithm parameters do not constitute a different method.
3. Re-submissions for a given method may be sent a maximum of once per week during the 6 month competition.
4. Since realistic future observations would include some low noise imaging, participants are welcome to use the GREAT08 Low-noise images to inform their GREAT08 Main analysis. We will never have observations for which the true shear is known, but we will be able to make our own attempts to simulate the sky, which could be used to train shear estimation methods. Therefore GREAT08 LowNoise-Known and GREAT08 RealNoise-Known have slightly different galaxy properties than GREAT08 LowNoise-Blind and GREAT08 RealNoise-Blind. GREAT08 LowNoise-Known and GREAT08 RealNoise-Known may be used to train the results of GREAT08 LowNoise-Blind and GREAT08 RealNoise-Blind.
5. Participants must provide a report detailing the method used, at the Challenge deadline. We would prefer that the code is made public.
6. We expect all participants to allow their results to be included in the final Challenge Report. We will however be flexible in cases where methods performed badly compared to the current methods if participants are strongly against publicising them.

We will release the true shears after the deadline and you are encouraged to write research articles using the Challenge simulations.

Some additional competition rules apply to members of the GREAT08 Team who submit entries:

7. For the purpose of these rules, "GREAT08 Team" includes anyone who receives STEP and/or GREAT08 Team emails, and/or has the STEP password. The authors of this document all receive GREAT08 Team emails.
8. Only information available to non-GREAT08 participants may be used in carrying out the analysis, for example, no inside information about the setup of the simulations may be used.

Note that the true blind shear values will only be available to only a small subset of the GREAT08 Team.

## APPENDIX B: OVERVIEW OF EXISTING METHODS

A variety of shear measurement methods have been developed by the cosmic lensing community. Their goal is always to obtain an unbiased estimate $\tilde{g}$ of the shear, such that the mean over a large population of galaxies is equal to the true shear $\langle\tilde{g}\rangle = g$. However, they adopt different approaches to correct the nuisance factors in Figure 4 (convolution, pixelisation and noise).

Most of the methods have been described, and tested on simulated images, during the Shear TEsting Programme (STEP) [Heymans et al. (2006), Massey et al. (2007a)] and earlier [Bacon et al. (2001), Erben et al. (2001), Hoekstra et al. (2002)]. To summarise the current level of knowledge, but trying not to restrict the development of new ideas, we present here an overview of an idealised method. In Appendices C–G, we then provide a more detailed introduction to several methods that have been used on real astronomical data, with links to research papers. At the launch of the GREAT08 Challenge, code for these methods will be made available and the corresponding results will be entered on the GREAT08 leader board.

Potential participants may be interested in applying methods that require a set of training data which matches the Challenge data. We do not provide such a set because this will not be available for realistic observations. It would in principle be possible to simulate data with similar properties to the observed data, but this will not match exactly because of our lack of knowledge of the detailed shapes of distant galaxies. We do not know whether or not this presents a fundamental limitation for this type of method. The (public) STEP1 and STEP2 simulations have a similar noise level to the GREAT08 images and the true shear is given. You are allowed to use these to train your methods if you wish. The galaxy properties are not the same as those in GREAT08 so this is a reasonable approximation to the realistic situation. However the objects are not isolated on postage stamps as for GREAT08.

**B.1. Ellipticity measurement.** We first describe a simple shear measurement method that would work in the absence of pixelisation, convolution and noise. The centre of the image brightness $I(x,y)$ can be defined via its first moments

$$\bar{x} = \frac{\int I(x,y) x \, dx \, dy}{\int I(x,y) \, dx \, dy}, \tag{B.1}$$

$$\bar{y} = \frac{\int I(x,y) y \, dx \, dy}{\int I(x,y) \, dx \, dy}, \tag{B.2}$$

and we can then measure the quadrupole moments

$$\mathcal{Q}_{xx} = \frac{\int I(x,y)(x-\bar{x})^2 \, dx \, dy}{\int I(x,y) \, dx \, dy}, \tag{B.3}$$

$$\mathcal{Q}_{xy} = \frac{\int I(x,y)(x-\bar{x})(y-\bar{y}) \, dx \, dy}{\int I(x,y) \, dx \, dy}, \tag{B.4}$$

$$\mathcal{Q}_{yy} = \frac{\int I(x,y)(y-\bar{y})^2 \, dx \, dy}{\int I(x,y) \, dx \, dy}. \tag{B.5}$$

Gravitational lensing maps the unlensed image, specified by coordinates $(x_{\rm u}, y_{\rm u})$, to the lensed image $(x_{\rm l}, y_{\rm l})$ using a matrix transformation

$$\begin{pmatrix} x_{\rm u} \\ y_{\rm u} \end{pmatrix} = \mathcal{A} \begin{pmatrix} x_{\rm l} \\ y_{\rm l} \end{pmatrix}, \tag{B.6}$$

where

$$\mathcal{A} = \begin{pmatrix} 1-g_1 & -g_2 \\ -g_2 & 1+g_1 \end{pmatrix}. \tag{B.7}$$

Throughout GREAT08, the components of shear $g_1$ and $g_2$ are constant across the image of a galaxy; this is usually a good approximation in real images too. Under this coordinate transformation, it can be shown that quadrupole moment tensor $\mathcal{Q}$ transforms as

$$\mathcal{Q}^{\rm u} = \mathcal{A} \mathcal{Q}^{\rm l} \mathcal{A}^T, \tag{B.8}$$

where $\mathcal{Q}^{\rm u}$ is the quadrupole moment tensor before lensing and $\mathcal{Q}^{\rm l}$ is that after lensing.

The overall ellipticity of a galaxy image can be quantified by the useful combination of moments [Bonnet and Mellier (1995)]

$$\epsilon \equiv \epsilon_1 + i\epsilon_2 = \frac{\mathcal{Q}_{xx} - \mathcal{Q}_{yy} + 2i\mathcal{Q}_{xy}}{\mathcal{Q}_{xx} + \mathcal{Q}_{yy} + 2(\mathcal{Q}_{xx}\mathcal{Q}_{yy} - \mathcal{Q}_{xy}^2)^{1/2}}, \tag{B.9}$$

where we introduce the complex notation $\epsilon = \epsilon_1 + i\epsilon_2$ and $g = g_1 + ig_2$ where $i^2 = -1$. For a simple galaxy that has concentric, elliptical isophotes (contours of constant brightness) with major axis $a$ and minor axis $b$, and angle

$\theta$ between the positive $x$ axis and the major axis,

$$\epsilon_1 = \frac{a-b}{a+b}\cos(2\theta), \tag{B.10}$$

$$\epsilon_2 = \frac{a-b}{a+b}\sin(2\theta). \tag{B.11}$$

The quantity $\epsilon$ transforms under shear as

$$\epsilon^{\mathrm{l}} = \frac{\epsilon^{\mathrm{u}} + g}{1 + g^*\epsilon^{\mathrm{u}}} \tag{B.12}$$

for $|g| < 1$, where the asterisk denotes complex conjugation [Seitz and Schneider (1997)]. This can be Taylor expanded to first order in $g$, for each of the two components $i \in 1, 2$.

To obtain measurements of $g$, we next assume that there is no preferred orientation for the shapes of galaxies in the absence of lensing. In this case, when averaged over a large population of galaxies, $\langle \epsilon_1^{\mathrm{u}} \rangle = \langle \epsilon_2^{\mathrm{u}} \rangle = 0$, $\langle \epsilon_1^{\mathrm{u}2} \rangle = \langle \epsilon_2^{\mathrm{u}2} \rangle$ and $\langle \epsilon_1^{\mathrm{u}} \epsilon_2^{\mathrm{u}} \rangle = 0$. Therefore, on Taylor expanding (B.10) to first order in $g$, we see that $\epsilon_i^{\mathrm{l}}$ is roughly a very noisy estimate of $g_i$ since $\sqrt{\langle \epsilon_i^{\mathrm{u}2} \rangle} \sim 0.15$, which is an order of magnitude larger than the typical value of $g_i$. On applying the symmetries for a large population we find

$$\langle \epsilon^{\mathrm{l}} \rangle \simeq g. \tag{B.13}$$

The need to sample a population of galaxies also explains the use of complex notation for both $\epsilon$ and $g$: the two components of $\epsilon$ average cleanly to zero in the absence of cosmic lensing, unlike a notation involving magnitude and angle. See Figure 5 for a graphical representation of these parameters.

More commonly considered is the combination of quadrupole moments

$$\chi = \frac{\mathcal{Q}_{xx} - \mathcal{Q}_{yy} + 2i\mathcal{Q}_{xy}}{\mathcal{Q}_{xx} + \mathcal{Q}_{yy}} \tag{B.14}$$

(sometimes known as "polarisation"), where we define components $\chi = \chi_1 + i\chi_2$ as before. This combination is more stable than $\epsilon$ in the presence of noise. A purely elliptical shape has

$$\chi_1 = \frac{a^2 - b^2}{a^2 + b^2}\cos(2\theta), \tag{B.15}$$

$$\chi_2 = \frac{a^2 - b^2}{a^2 + b^2}\sin(2\theta). \tag{B.16}$$

In general, $\chi$ transforms under shear as

$$\chi^{\mathrm{l}} = \frac{\chi^{\mathrm{u}} + 2g + g^2\chi^{\mathrm{u}*}}{1 + |g|^2 + 2\Re(g\chi^{\mathrm{u}*})}, \tag{B.17}$$

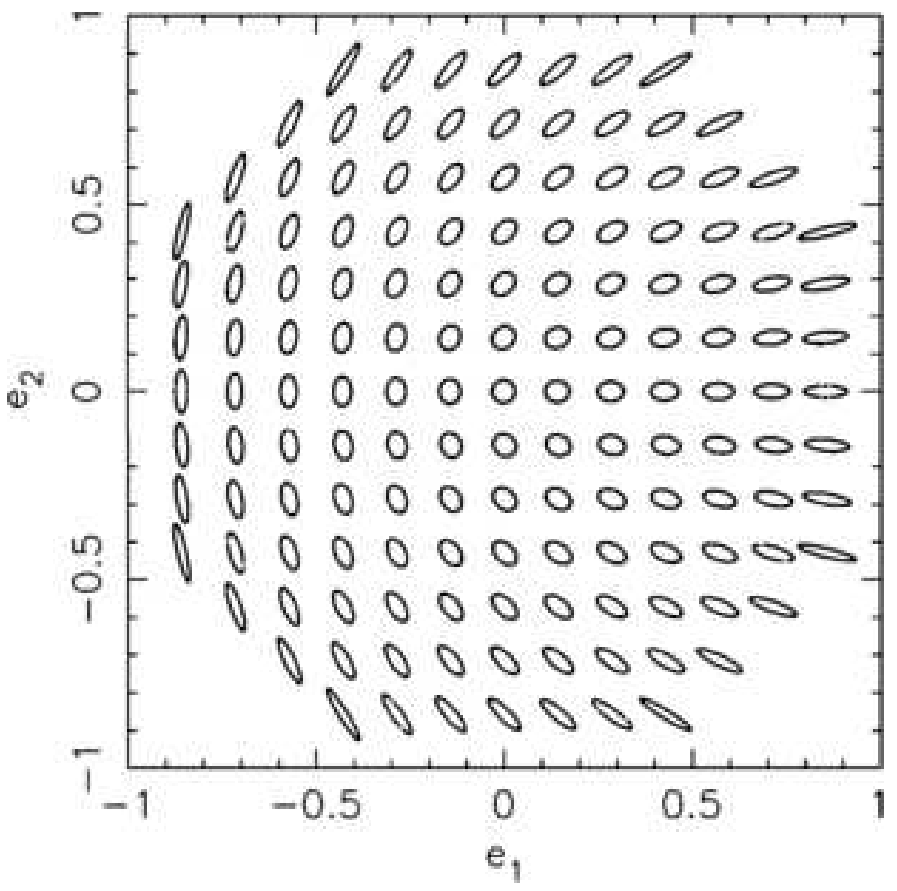

Fig. 5. *The parameterisation of generalised ellipticity as two quantities $e_1$ and $e_2$, showing the shape of isophotes of an elliptical galaxy. For example, a galaxy aligned with the $y$ axis ($\theta = 90$ degrees) has $e_1 < 0$ and $e_2 = 0$. If this figure had shown individual ellipticity estimates like $\epsilon$ or $\chi$, the orientations would be the same, but the elongations would vary.*

where $\Re$ denotes that the real part should be taken [Schneider and Seitz (1995)]. On Taylor expanding to first order in $g$ and averaging over a population for which $\langle\chi_1^{\rm u}\rangle = \langle\chi_2^{\rm u}\rangle = 0$, $\langle\chi_1^{\rm u2}\rangle = \langle\chi_2^{\rm u2}\rangle$ and $\langle\chi_1^{\rm u}\chi_2^{\rm u}\rangle = 0$ we obtain

$$\langle\chi^{\rm l}\rangle \simeq 2(1 - \langle\chi_1^{\rm u2}\rangle)g. \tag{B.18}$$

Therefore if the variance of the unlensed ellipticities $\langle\chi_1^{\rm u2}\rangle$ of the population is known then the shear can be approximately determined. For GREAT08 it may be possible to infer these ellipticity properties from the low noise sample since $\langle\chi_i^{\rm u2}\rangle \simeq \langle\chi_i^{\rm l2}\rangle$. For more information, see Section 4 of Bartelmann and Schneider (2001) or Bernstein and Jarvis (2002).

These are just two examples of generalised ellipticity estimates for a galaxy shape. All existing methods start in similar fashion, by constructing a mapping from the 2D image $I(x, y)$ to a quantity with the rotational symmetries of an ellipse, such as $\epsilon$ or $\chi$. For some methods, the mapping might involve a combination of quadrupole moments. To reduce contamination from neighbouring galaxies, and to limit the impact of noise in the wings of a galaxy, a weight function $W(x, y)$ with finite support is normally included in equations (B.3), (B.4) and (B.5). Other methods might involve the fitting of a parametric (e.g., elliptical Gaussian, or exponential) model to the image, in which case the major and minor axes $a$ and $b$ are returned, along with the angle $\theta$. Various basis functions have been tried for this modelling, including shapelets (Appendix D), sums of co-elliptical Gaussians (see Appendices E and G) de Vaucouleurs profiles (Appendix F). Each support a different range of potential galaxy shapes, and have had varying success on galaxies of different morphological type.

**B.2. Shear responsivity.** Converting a general ellipticity measurement $e$ into a shear estimate $\tilde{g}$ also requires knowledge of how that ellipticity is affected by a shear. All existing shear measurement methods involve some form of ellipticity estimate and corresponding shear responsivity

$$P_{ij}^{\mathrm{sh}} = \frac{\partial e_i}{\partial g_j} \tag{B.19}$$

(sometimes called the shear polarisability or susceptibility) so that

$$e_i^{\mathrm{l}} = e_i^{\mathrm{u}} + P_{ij}^{\mathrm{sh}} g_j + \mathcal{O}(g^2), \tag{B.20}$$

where $j$ is summed over. In general, $P_{ij}^{\mathrm{sh}}$ is a unique $2 \times 2$ tensor for each galaxy. The diagonal elements reflect how much a shear in one direction alters the ellipticity in the same direction, and the two diagonal elements tend to be similar. The off-diagonal elements reflect the degree to which a shear in one direction alters the ellipticity in the other, and tend to be small. For the present purposes, it is therefore reasonable to think of the shear responsivity for each ellipticity estimate as a scalar quantity $P^{\mathrm{sh}}$ times the identity matrix. Expressions for $P^{\mathrm{sh}}$ for three simple ellipticity measures are shown in Table 3. In general, shear responsivity depends on the ellipticity and cuspiness of an individual galaxy image and can even depend on the shear. For example, the axis ratio of a circle initially changes significantly under a small shear operation; but as the same shear is repeatedly reapplied, the object can tend toward a straight line but then its ellipticity can never increase further since $|e| < 1$. A shear estimate can then be formed via

$$\tilde{g}_i \equiv \frac{e_i}{P^{\mathrm{sh}}}. \tag{B.21}$$

If we had access to the noise-free, unlensed galaxy image then we could calculate $P^{\mathrm{sh}}$ for each galaxy. However the lensing signal does not change over time, and the strongest cosmic lensing signal is carried by the most distant—and therefore the faintest—galaxies. Measurements of $P^{\rm sh}$ from the observed image are consequently very noisy. Since $P^{\rm sh}$ is on the denominator of equation (B.21), errors in this quantity can contribute to potential biases and large wings in the global distribution of $\tilde{g}$. Getting it wrong in existing methods has typically led to a bias in shear measurement that is proportional to the shear ("multiplicative bias"). To reduce the noise and control bias, $P^{\rm sh}$ is often averaged over or fitted from a large population of galaxies. It is typically fitted as a function of galaxy size and brightness (the distribution of true galaxy shapes is known to vary as a function of these observables). However, the fitting function must be chosen carefully: shear responsivity often varies rapidly as a function of galaxy brightness, and existing methods have been found to be unstable with respect to the method used for this fitting. Sometimes $P^{\rm sh}$ is also fitted as a function of ellipticity. This drastically overestimates the cosmological shear signal in intrinsically elliptical galaxies, but this should average out over a population. The goal is merely to create a shear estimate that is unbiased for a large population.

TABLE 3
*Some common ellipticity estimates and their corresponding shear responsivities, calculated to first order in $g$*

| **Ellipticity estimate** | **Shear responsivity** |
|---|---|
| $\epsilon = (\frac{a-b}{a+b})(\cos(2\theta) + i\sin(2\theta))$ | 1 |
| $\chi = (\frac{a^2-b^2}{a^2+b^2})(\cos(2\theta) + i\sin(2\theta))$ | $2(1 - \langle \chi_1^{\mathrm{u}2} \rangle)$ |
| $\frac{\int I(x,y)(x^2-y^2+2ixy)W(x,y)\,dx\,dy}{\int I(x,y)(x^2+y^2)W(x,y)\,dx\,dy}$ | Eqn. (5-2) in Kaiser, Squires and Broadhurst (1995) |

Shear responsivity thus represents the intrinsic morphology of an individual galaxy, or the morphology distribution for a population of galaxies. Although inferring the intrinsic shape distribution is not a goal in itself (see Figures 2 and 3), some aspect of it always needs to be measured. As discussed in Appendix F, it arises in a Bayesian context as a prior on probability distribution for each shear estimate.

**B.3. Correcting for a convolution kernel.** An image is inevitably blurred by a convolution kernel (generally known in astronomy literature as the Point Spread Function or PSF) introduced by the camera optics and atmospheric turbulence. The kernel is usually fairly compact, and two examples are given in Figure 6. The typical size is usually quantified by the Full Width Half Maximum (FWHM), which is the diameter where the light falls to half of the peak. Typically the FWHM is two or three pixels across, and of a similar size to the galaxies of interest. For a Gaussian kernel the Gaussian standard deviation is simply related to the FWHM via $\mathrm{FWHM} = 2\sqrt{2\ln(2)}\sigma$.

One approach to correct for the kernel, which is particularly useful for moment-based ellipticities, is to subtract the effects of the convolution kernel from both the ellipticity and the shear responsivity [e.g., equations (C.5) and (C.6)]. A second approach to correct for the convolution kernel, particularly appropriate for fitting methods, has been a full deconvolution of the image. One fairly stable way to do this has been the forward convolution of a predefined set of basis functions with a model of the convolution kernel, followed by the fitting of these basis functions to the data. A deconvolved version of the image can then be reconstructed by using the derived model coefficients with the (unconvolved) basis functions. This model can then be used to measure an ellipticity and shear responsivity. Getting this step wrong in existing methods can leave residual effects of (or overcorrects for) any anisotropy of the convolution kernel in the ellipticity estimate. This typically introduces a bias in shear measurement that is independent of the true shear ("additive bias").

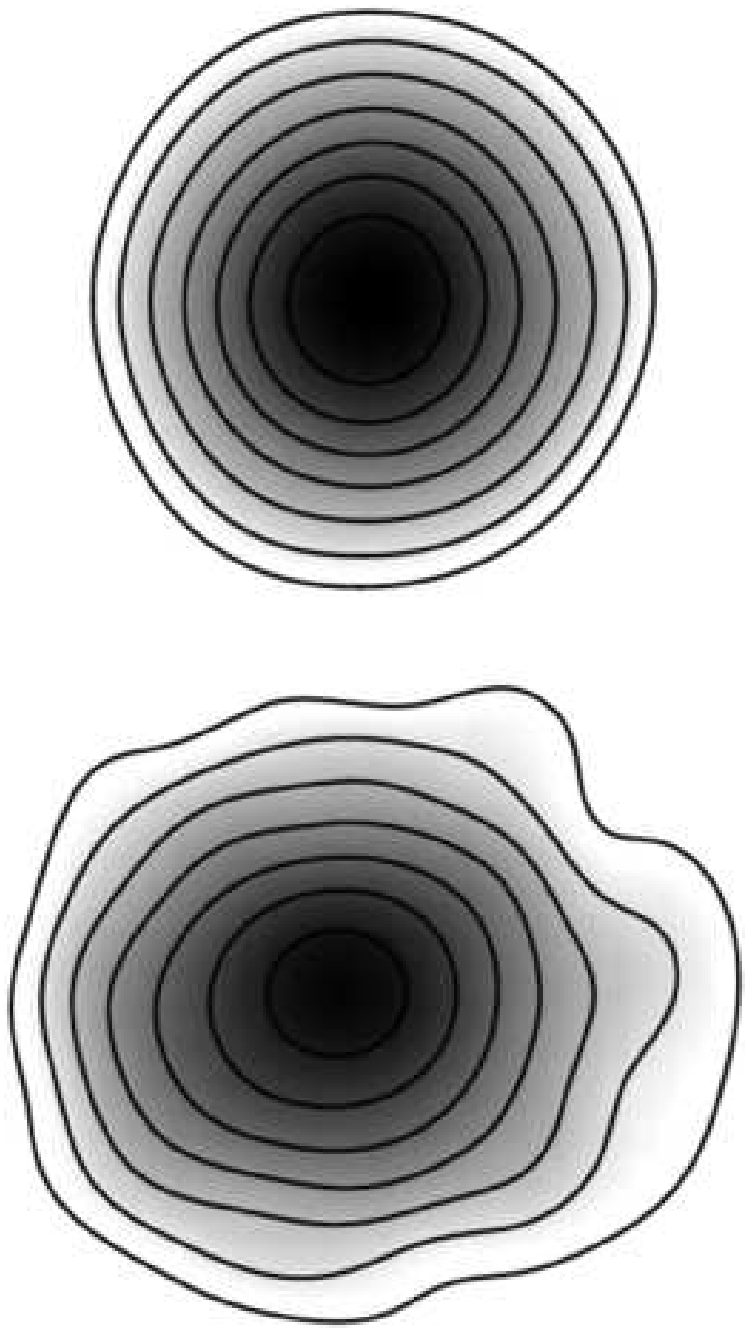

Fig. 6. *Detail of two realistic convolution kernels. The isophotal contours are logarithmically spaced.*

**B.4. Correcting for pixelisation.** Astronomical detectors for optical light count the total number of photons arriving in a region that we call a pixel. To a good approximation these pixels are on a square grid and do not overlap or have gaps between them. For methods that fit a model to each galaxy shape, including a forward convolution with the convolution kernel, pixelisation can in principle be incorporated easily. This is because integration within a square pixel is mathematically identical to convolution with a square top hat, followed by resampling at the centres of pixels. Since the observed images of stars have also been pixelised, they are already a rendering of the convolution kernel convolved with the square of the pixel. Deconvolving this naturally takes care of the pixelisation at the same time. In practice, models for the kernel are relatively smooth and may not capture the convolution with the square well.

No methods based on quadrupole moments, with correction for the convolution kernel via subtraction of those moments, have yet included a proper treatment of pixelisation. Furthermore, for both types of method, an unexplained difference has been observed [Massey et al. (2007d)]. This is particularly important to the cosmic lensing community because the design of some future telescopes currently feature only about 2 pixels across the FWHM of the convolution kernel.

**B.5. Averaging to remove noise.** There are two contributions to the noise on a shear estimate $\tilde{g}$ for a single galaxy. The first comes from the noise on the image, which is Poisson for GREAT08. The second comes from the fact that unlensed galaxies are not circular and thus it is not possible to tell for a single galaxy whether it is intrinsically elliptical or whether it is intrinsically circular and lensed by a strong shear. This can be beaten down by averaging the ellipticities of many galaxies. If the galaxies are in a similar location (or within the same set of GREAT08 images), the constant shear signal they contain will be all that remains. Unfortunately, almost all existing shear measurement methods supply only a single (maximum likelihood) shear estimate for each galaxy, possibly with a single error bar (although see Appendices E and F). The PDF is not exactly a Gaussian, therefore a simple average is not the correct approach.

## APPENDIX C: EXISTING METHOD 1: WEIGHTED QUADRUPOLE MOMENTS (KSB+)

Currently, the most widely used and oldest method for cosmic lensing analysis comes from the work of Kaiser, Squires and Broadhurst (1995), Luppino and Kaiser (1997) and Hoekstra et al. (1998), hereafter referred to as KSB+. The version of KSB+ made available for the GREAT08 challenge is the "CH" KSB pipeline documented in the STEP challenge [Heymans et al. (2006) and Heymans et al. (2005)]. The original KSB imcat software developed by Nick Kaiser is also available on request.

KSB+ parameterises galaxies and stars according to their weighted quadrupole moments

$$\mathcal{Q}^{\mathrm{w}}_{ij} = \frac{\int I(x,y) x_i x_j W(x,y)\,dx\,dy}{\int I(x,y) W(x,y)\,dx\,dy}, \tag{C.1}$$

where $W$ is a Gaussian weight function of scale length $r_{\mathrm{g}}$, where $r_{\mathrm{g}}$ is some measure of galaxy size such as the half-light radius and $x_1 = x - \bar{x}$, $x_2 = y - \bar{y}$. An ellipticity $\varepsilon$ is formed from these weighted moments using equation (B.14). The following KSB+ method details how to correct for the convolution kernel and get an unbiased estimate of the shear $\gamma$.

The main limiting simplification in KSB+ is to assume that the convolution kernel can be described as a small but highly anisotropic distortion convolved with a large circularly symmetric function. In most instances, this is not a good approximation to make, but the KSB+ method has proved to be remarkably accurate in practice. With this assumption, the "corrected ellipticity" of a galaxy (which it would have in perfect observations) $\varepsilon^{\mathrm{cor}}$, is given by

$$\varepsilon^{\mathrm{cor}}_{\alpha} = \varepsilon^{\mathrm{obs}}_{\alpha} - P^{\mathrm{sm}}_{\alpha\beta} p_{\beta}, \tag{C.2}$$

where $p$ is a vector that measures the kernel anisotropy, and $P^{\mathrm{sm}}$ is the smear responsivity tensor given in Hoekstra et al. (1998). The kernel anisotropy $p$ can be estimated from images of stellar objects by noting that a star, denoted by an asterisk, has zero ellipticity (it is effectively a $\delta$-function) before convolution: $\varepsilon^{*\mathrm{cor}}_{\alpha} = 0$. Hence,

$$p_{\mu} = (P^{\mathrm{sm}*})^{-1}_{\mu\alpha} \varepsilon^{*\mathrm{obs}}_{\alpha}. \tag{C.3}$$

The isotropic effect of the convolution kernel and the smoothing effect of the weight function $W$, can be accounted for by applying a tensor correction $P^{\gamma}$, such that

$$\varepsilon^{\mathrm{cor}}_{\alpha} = \varepsilon^{\mathrm{s}}_{\alpha} + P^{\gamma}_{\alpha\beta} \gamma_{\beta}, \tag{C.4}$$

where $\varepsilon^{\mathrm{s}}$ is the intrinsic source ellipticity and $\gamma$ is the gravitational shear. Luppino and Kaiser (1997) show that

$$P^{\gamma}_{\alpha\beta} = P^{\mathrm{sh}}_{\alpha\beta} - P^{\mathrm{sm}}_{\alpha\mu} (P^{\mathrm{sm}*})^{-1}_{\mu\delta} P^{\mathrm{sh}*}_{\delta\beta}, \tag{C.5}$$

where $P^{\mathrm{sh}}$ is the shear responsivity tensor given in Hoekstra et al. (1998) and $P^{\mathrm{sm}*}$ and $P^{\mathrm{sh}*}$ are the stellar smear and shear responsivity tensors, respectively. Combining the correction for the anisotropic part of the convolution kernel [equation (C.4)] and the $P^{\gamma}$ isotropic correction, the final KSB+ shear estimate $\hat{\gamma}$ is given by

$$\hat{\gamma}_{\alpha} = (P^{\gamma})^{-1}_{\alpha\beta} [\varepsilon^{\mathrm{obs}}_{\beta} - P^{\mathrm{sm}}_{\beta\mu} p_{\mu}]. \tag{C.6}$$

This method has been used by many astronomers although different interpretations of the above formula have introduced some subtle differences between each astronomer's KSB+ implementation.

Other methods inspired by KSB+ can be found in Hirata and Seljak (2003), Mandelbaum et al. (2005), Rhodes, Refregier and Groth (2000), Kaiser (2000) and Smith et al. (2001).

## APPENDIX D: EXISTING METHOD 2: SHAPELETS

An orthonormal basis set, referred to as "shapelets," can be formed by the product of Gaussians with Hermite or Laguerre polynomials (in Cartesian or polar coordinates respectively). A weighted linear sum of these basis functions can model any compact image, including the irregular spiral arms and bulges seen in galaxy shapes [Refregier (2003a), Massey and Refregier (2005)]. The shapelet transform acts qualitatively like a localised Fourier transform, and can be used to filter out high frequency features such as noise.

The shapelet basis functions are not specifically optimised for the compression of galaxy shapes. However, they can be analytically integrated within pixels and have particularly elegant and convenient expressions for convolution and shear operations. After modelling both a galaxy shape and a convolution kernel as a linear combination of shapelet basis functions, convolution can be expressed as a simple matrix multiplication [see also Berry, Hobson and Withington (2004)]. Deconvolution can be performed via a matrix inversion, although in practice appears more stable when performed via a forward convolution of the basis functions, then obtaining their coefficients with a fast, least-squares fit. Shearing a shapelet model involves mixing between only a minimal number of model coefficients.

Most of the parameters in a shapelet model are linear, which helps minimise any potential biases that could arise when fitting faint, noisy images. Additional, nonlinear parameters are the overall scale size and the coordinates to the centre of the basis functions, plus the finite truncation order of the shapelet series. Each fitted nonlinear parameter requires a slower, nonlinear iteration to pre-defined goals. Some methods also use elliptical shapelet basis functions, derived by shearing circular shapelets: such methods require two additional nonlinear parameters (the two ellipticity components).

Shapelet basis functions have been utilised in various ways, for both iterative and noniterative shear measurement methods. There are three approaches currently in the literature:

- The shapelet modelling process is used to obtain a best-fit denoised, deconvolved and depixelised image from which quadrupole moments are calculated. Experiments with various functional forms for the radial shape of the weight function have been tried in Refregier and Bacon (2003) and Massey and Refregier (2005). Different weight functions provide a variety of benefits, primarily altering the shear responsivity factor (B.19).
- A perfectly circular model galaxy with arbitrary radial profile is sheared and convolved until it best matches the observed image according to a least-squares criterion [Kuijken (2006)]. A subset of the shapelet basis is used as a way of allowing freedom in the radial profile. The probability distribution function of galaxy ellipticities is required, in order to calibrate how much of the shearing is required to account for intrinsic shapes.
- A shapelet model for the galaxy is constructed which is "circular" by a particular definition. Unlike the previous bullet point, it need not be circularly symmetric, but is constrained to have zero ellipticity for a particular ellipticity definition. This is then sheared and convolved until it matches the data. This is discussed by Bernstein and Jarvis (2002) and tested by Nakajima and Bernstein (2007). This similarly requires the probability distribution function of intrinsic galaxy ellipticities.

Concerns have been raised that the Gaussian-based functions require a large number of coefficients to reproduce the extended, low-level wings of typical galaxies. If these wings are hidden beneath noise, and truncated in the model, the galaxy's ellipticity will be systematically underestimated. Initial experiments are attempting to replace the Gaussian part of shapelets with something better matched to galaxy shapes, like a sech or an exponential (Kuijken, in prep.). Appropriate polynomials can always be used to generate an orthonormal basis set, and this should extrapolate better into the wings. It might be possible to transfer experience with Gaussian shapelets to these new basis sets. The elegant image manipulation operations would made significantly more complicated, and involve mixing between many, nonneighbouring coefficients. However, the mixing matrices can still be pre-calculated for a given basis set as a look-up table.

More information, links to the papers, and a software package for shapelet modelling in the IDL language can be obtained from http://www.astro.caltech.edu/~rjm/shapelets. Translations of the code into C++ and java may also be available upon request.

## APPENDIX E: EXISTING METHOD 3: FITTING SUMS OF CO-ELLIPTICAL GAUSSIANS

Kuijken (1999) presented a maximum likelihood method in which each galaxy and convolution kernel is modelled as a sum of elliptical Gaussians. The implementation below follows Bridle et al. (2002) (im2shape) and Voigt and Bridle (2008). The model intensity $B(\mathbf{x})$ as a function of position $\mathbf{x} = (x, y)$ is

$$B(\mathbf{x}) = \sum_i \frac{A_i}{2\pi|C_i|^{-1/2}} e^{-(\mathbf{x}-\mathbf{x}_i)^T C_i (\mathbf{x}-\mathbf{x}_i)/2}, \tag{E.1}$$

where the inverse covariance matrix for each component $C_i$ can be written in terms of the ellipse major and minor axes ($a_i$ and $b_i$) as

$$(C_i)_{1,1} = 2\left(\frac{\cos^2(\theta_i)}{a_i^2} + \frac{\sin^2(\theta_i)}{b_i^2}\right), \tag{E.2}$$

$$(C_i)_{1,2} = \left(\frac{1}{b_i^2} - \frac{1}{a_i^2}\right)\sin(2\theta_i), \tag{E.3}$$

$$(C_i)_{2,2} = 2\left(\frac{\cos^2(\theta_i)}{b_i^2} + \frac{\sin^2(\theta_i)}{a_i^2}\right) \tag{E.4}$$

and the matrix is symmetric. Thus each Gaussian object component has 6 parameters, which we consider to be the position of the centre $\mathbf{x}_i = (x_i, y_i)$, $|\epsilon_i| \equiv (a_i - b_i)/(a_i + b_i)$, $\theta_i$, $r_i \equiv a_i b_i$ and the amplitude $A_i$. Because the galaxy is a sum of Gaussians, convolution with the convolution kernel (another sum of Gaussians) is analytically simple.

The likelihood of the parameters is calculated assuming that the noise on the image is Gaussian with unknown variance $\sigma$ and that an unknown constant background level $b$ has been added to the image. The model parameter vector $\mathbf{p}$ thus consists of $\mathbf{p} = (\sigma, b, x_1, y_1, |\epsilon_1|, \theta_1, ab_1, A_1, \ldots, x_n, y_n, e_n, \theta_n, ab_n, A_n)$, where the subscripts denote the Gaussian component number and $n$ is the number of Gaussian components that make up the object. To reduce the number of parameters, the centre position, ellipticity and angle of all components in each galaxy are fixed to be the same. Thus each additional Gaussian contributes only two extra parameters. This is a significant limitation on the flexibility of the galaxy model, but makes the method more stable to noise in the image, and means that the shear estimate is equal to the ellipticity $\epsilon$ of the Gaussian stack via equation (B.13). This scheme will not accurately model irregular galaxy shapes, but that is not the main goal.

Each parameter in $p$ is assigned a prior which allows the conversion to the posterior probability $\mathcal{P}(\mathbf{p}|\mathbf{D}, \mathrm{PSF})$, assuming that the convolution kernel (PSF) is known exactly. Markov-chain Monte Carlo sampling is used to find the marginalised PDF in $\epsilon_1$, $\epsilon_2$ space. This must be combined with the PDF of unlensed galaxy ellipticities to find the PDF in $g_1$, $g_2$ space. In practice the mean and standard deviation of the samples in $\epsilon_1$ and $\epsilon_2$ space are calculated. These are converted to shear estimates by adding the root mean square of the unlensed ellipticities $\langle(\epsilon_i^{\mathrm{u}})^2\rangle$ in quadrature with the standard deviation of the samples.

## APPENDIX F: EXISTING METHOD 4: LENSFIT—BAYESIAN SHEAR ESTIMATE WITH REALISTIC GALAXY MODEL FITTING

Lensfit is a model fitting shape measurement method that uses a Bayesian shear estimate to remove biases. A Bayesian estimate has the immediate advantage over likelihood based techniques in that, as described in Miller et al. (2007), due to the inclusion of a prior the shear estimate should be unbiased given an ideal shape measurement method and an accurate prior. Miller et al. (2007) also discuss how to remove any bias that occurs as a result of assuming that the prior is centred on zero ellipticity, which is assumed since the actual intrinsic distribution is unknown.

For each galaxy a (Bayesian) posterior probability in ellipticity can be generated using

$$p_i(\mathbf{e}|\mathbf{y}_i) = \frac{\mathcal{P}(\mathbf{e})\mathcal{L}(\mathbf{y}_i|\mathbf{e})}{\int \mathcal{P}(\mathbf{e})\mathcal{L}(\mathbf{y}_i|\mathbf{e})\, d\mathbf{e}}, \tag{F.1}$$

where $\mathcal{P}(\mathbf{e})$ is the ellipticity prior probability distribution and $\mathcal{L}(\mathbf{y}_i|\mathbf{e})$ is the likelihood of obtaining the $i$th set of data values $\mathbf{y}_i$ given an intrinsic ellipticity (i.e., in the absence of lensing) $\mathbf{e}$. By considering the summation over the data, the true distribution of intrinsic ellipticities can be obtained from the data itself

$$\left\langle \frac{1}{N}\sum_i p_i(\mathbf{e}|\mathbf{y}_i)\right\rangle = \int d\mathbf{y}\, \frac{\mathcal{P}(\mathbf{e})\mathcal{L}(\mathbf{y}|\mathbf{e})}{\int \mathcal{P}(\mathbf{e})\mathcal{L}(\mathbf{y}|\mathbf{e})\, d\mathbf{e}} \int f(\mathbf{e})\epsilon(\mathbf{y}|\mathbf{e})\, d\mathbf{e}, \tag{F.2}$$

where, on the right-hand side, the integration of the probability distribution gives the expectation value of the summed posterior probability distribution for the sample. $\epsilon(\mathbf{y}|\mathbf{e})$ is the probability distribution for $\mathbf{y}$ given $\mathbf{e}$. This will yield the true intrinsic distribution under the conditions that $\epsilon(\mathbf{y}|\mathbf{e}) = \mathcal{L}(\mathbf{y}|\mathbf{e})$ and $\mathcal{P}(\mathbf{e}) = f(\mathbf{e})$ (assuming the likelihood is normalised) from which we obtain

$$\left\langle \frac{1}{N}\sum_i p_i(\mathbf{e}|\mathbf{y})\right\rangle = \mathcal{P}(\mathbf{e}) = f(\mathbf{e}). \tag{F.3}$$

This is the equation that highlights the essence of the Bayesian shape measurement method, given a prior that is a good representation of the intrinsic distribution of ellipticities the estimated posterior probability should be unbiased. Kitching et al. (2008) discuss how to find the prior from a subset of the data itself. The shear is equal to the average expectation value of the ellipticity with a factor $\partial\langle\mathbf{e}\rangle_i/\partial\mathbf{g}$ which corrects for any incorrect assumptions about the prior

$$\tilde{\mathbf{g}} = \frac{\sum_i^N \langle\mathbf{e}\rangle_i}{\sum_i^N |\partial\langle\mathbf{e}\rangle_i/\partial\mathbf{g}|}, \tag{F.4}$$

where for an individual galaxy the $\langle e\rangle = \int eP(\mathbf{e})\, de$. The shear responsivity is calculated by finding the derivative of ellipticity with respect to the shear. Miller et al. (2007) show how this can be calculated directly from the prior and the likelihood in a Bayesian shear estimation method.

To generate the full likelihood surface in $(e_1, e_2)$, we fit a de Vaucouleurs profile to each galaxy image. This results in six free parameters per galaxy: position $x$, position $y$, $e_1$, $e_2$, brightness and a scale factor $r$. By doing the model fitting in Fourier space the brightness and position can be marginalised over analytically, leaving the ellipticity and radius to fit. The radius is then numerically marginalised over leaving a likelihood as a function of ellipticity. This likelihood is then used in the Bayesian formalism above to estimate the shear.

## APPENDIX G: EXISTING METHOD 5: MODEL-FITTING METHOD WITH NONLINEAR DISTORTION TERMS

This model-fitting method goes beyond those in which distortion is entirely parameterised by the linear effect of shear. In addition to ellipticity, nonlinear shapes are measured by using generalised versions of transformation (2.1) that include second-order terms arising if the shear signal varies across the width of a galaxy (it does not in the GREAT08 simulations). The models simultaneously allow for the estimation of these nonlinear parameters, which should yield a more reliable estimation of shear, and are also of use in cosmology.

This method uses a compact form for the generalised transformations through the use of complex variables $\{w = x + iy, \bar{w} = x - iy\}$, where $\bar{w}$ is the complex conjugate of $w$. In this notation, equation (2.1) is simply written $w^{\mathrm{u}} = w - g\bar{w}^{\mathrm{l}}$, where the superscripts "u" and "l" refer to the unlensed and lensed images respectively. That transformation can be generalised to

$$w^{\mathrm{u}} = w^{\mathrm{l}} - g\bar{w}^{\mathrm{l}} - b\bar{w}^{\mathrm{l}2} - \bar{d}w^{\mathrm{l}2} - 2d\bar{w}^{\mathrm{l}}w^{\mathrm{l}}, \tag{G.1}$$

where additional nonlinear terms are introduced, with complex coefficients $\{b = b_1 + ib_2, d = d_1 + id_2\}$ and $\bar{d} = d_1 - id_2$. See Irwin and Shmakova (2005) and Schneider and Er (2007) for details.

This is a direct fitting method that uses an assumed model for a galaxy's radial profile $F(r)$. The radial position $r$ has a straightforward expression in the complex notation, with $r^2 = x^2 + y^2 = \bar{w}w$. The intensity of the model as a function of position $(x_{\mathrm{l}}, y_{\mathrm{l}})$ for a lensed galaxy will have a form

$$F(w^{\mathrm{u}} - (w^{\mathrm{u}})_0) \to F(w^{\mathrm{l}} - g\bar{w}^{\mathrm{l}} - b(\bar{w}^{\mathrm{l}})^2 - \bar{d}w^{\mathrm{l}2} - 2d\bar{w}^{\mathrm{l}}w^{\mathrm{l}} - (w^{\mathrm{l}})_0), \tag{G.2}$$

where $(w_i^{\mathrm{u}})_0$ is the centroid position. The function $F(r)$ could be any radial profile function: for example a Gaussian, sum of Gaussians, a Gaussian times a Polynomial, de Vaucouleurs, exponential or a parametric spline function. This function represents a galaxy model before convolution with a kernel. It is convolved with the convolution kernel and then fitted to the galaxy image.

Irwin and Shmakova (2005) and Irwin, Shmakova and Anderson (2007) used a Gaussian times a Polynomial profile as a model function

$$F(r^{\mathrm{u}2}) = (A + Br^{\mathrm{u}2} + Cr^{\mathrm{u}4}) + e^{-Dr^{\mathrm{u}2}}, \tag{G.3}$$

where $A$ is related to the intensity at the centre of the galaxy, $B$ is for a better fit to an arbitrary behaviour at the origin, $D$ is a cut-off scale that reflects the image size, and $C$ can modify the behaviour as one approaches the size of the image. The "+" subscript indicates that if the polynomial has a value less than zero, it is to be set equal to zero. This is needed to avoid negative intensities, which would be unphysical.

The parameters of the radial profile $\{A, B, C, D\}$, the shape transformation parameters $\{g, b, d\}$ and the centroid position $w_0$ are determined by minimizing the norm

$$\|I_F - I_{\rm l}\|_\omega^2 = \int (I_F - I_{\rm l})^2 \omega \, dx_{\rm l} \, dy_{\rm l}, \tag{G.4}$$

where $I_F$ is given by convolving $F(r^{\rm u2})$ with the PSF convolution kernel. In the model function $I_F(r^{\rm u2})$, $r^{\rm u2} = w^{\rm u}\bar{w}^{\rm u}$, is understood to be a function of $x^{\rm l}$ and $y^{\rm l}$ through $w^{\rm l}$ and $\bar{w}^{\rm l}$. A weight $\omega$ can be introduced to account for measurement uncertainty in each pixel if some are known to be more noisy than others.

With the extra parameters $b, d$ included in the shape distortion, as well as shear $g$, in addition to the radial shape parameters $\{A, B, C, D\}$ and the centroid position, $w_0 \to (x_{\rm l}, y_{\rm l})$, there are 12 variables to determine. The fit is done in several steps using a multi-dimensional Newton's method. At each step a subset of the 12 variables are allowed to vary. The curvature matrix for these parameters is computed then diagonalised, and eigenvectors with very small eigenvalues are not allowed to contribute to the function change at that step. The rate of convergence to a minimum is controlled by a parameter step size.

This method has an advantage over other methods in that the models can represent a better fit to a galaxy image for galaxies with nonelliptical isophotes. In addition one of the challenging tasks of ellipticity measurements is in the definition of a galaxy's centroid. The centroid position is affected by the nonlinear terms and the simultaneous definition of these parameters will give a better centroid measurement.

**Acknowledgements.** This project was born from the Shear TEsting Programme and a clinic at the University College London (UCL) Centre for Computational Statistics and Machine Learning (CSML). The GREAT08 Challenge is a Pattern Analysis, Statistical Modelling and Computational Learning (PASCAL) Challenge. PASCAL is a European Network of Excellence under Framework 6. We thank John Bridle, Michiel van de Panne, Michele Sebag, Antony Lewis, Christoph Lampert, Bernhard Schoelkopf, Chris Williams, David MacKay, Maneesh Sahani, David Barber and Nick Kaiser for helpful discussions.

S. Bridle
J. Shawe-Taylor
S. T. Balan
D. Kirk
D. Witherick
L. Voigt
Department of Physics and Astronomy
University College London
Gower Street
London, WC1E 6BT
United Kingdom
E-mail: sarah.bridle@ucl.ac.uk

A. Amara
University of Hong Kong
Pok Fu Lam Road
Hong Kong

D. Applegate
M. Shmakova
Stanford Linear Accelerator Center
National Accelerator Laboratory
2575 Sand Hill Road
Menlo Park, California 94025
USA

J. Berge
B. Moghaddam
J. Rhodes
Jet Propulsion Laboratory
4800 Oak Grove Drive
Pasadena, California 91109
USA

J. Berge
R. Massey
B. Moghaddam
J. Rhodes
California Institute of Technology
1200 East California Boulevard
Pasadena, California 91125
USA

J. Berge
Y. Moudden
S. Paulin-Henriksson
S. Pires
A. Rassat
A. Refregier
Commissariat a l'Energie Atomique, Saclay
Bat 709—Orme des Meurisiers
SAp—CEA/Saclay
91191 Gif sur Yvette Cedex
France

G. Bernstein
M. Jarvis
R. Nakajima
Department of Physics and Astronomy
University of Pennsylvania
209 South 33rd Street
Philadelphia, Pennsylvania 19104-6396
USA

H. Dahle
J.P. Kneib
Laboratoire d'Astrophysique de Marseille
Observatoire Astronomique de Marseille-Provence
Pôle de l'Étoile Site de Château-Gombert
38, rue Frédéric Joliot-Curie
13388 Marseille cedex 13
France

T. Erben
E. Semboloni
Argelander-Institut für Astronomie
University of Bonn
Auf dem Hügel 71
D-53121 Bonn
Germany

M. Gill
Department of Physics
The Ohio State University
191 West Woodruff Avenue
Columbus, Ohio 43210
USA

A. Heavens
Institute for Astronomy
Royal Observatory
University of Edinburgh
Blackford Hill
Edinburgh EH9 3HJ
United Kingdom

C. Heymans
L. van Waerbeke
University of British Columbia
2329 West Mall
Vancouver, BC V6T 1Z4
Canada

F. W. High
Harvard-Smithsonian Center for Astrophysics
Harvard University
60 Garden Street
Cambridge, Massachusetts 02138
USA

H. Hoekstra
University of Victoria
PO Box 1700 STN CSC
Victoria, BC V8W 2Y2
Canada

T. Kitching
Oxford Astrophysics
Department of Physics
University of Oxford
Denys Wilkinson Building
Keble Road
Oxford, OX1 3RH
United Kingdom

K. Kuijken
T. Schrabback
Leiden Observatory
Huygens Laboratory
University of Leiden
J. H. Oort Building
Niels Bohrweg 2
NL-2333 CA Leiden
The Netherlands

D. Lagatutta
D. Wittman
Physics Department
University of California, Davis
One Shields Avenue
Davis, California 95616
USA

R. Mandelbaum
Institute for Advanced Study
Einstein Drive
Princeton, New Jersey 08540
USA

C. Heymans
Y. Mellier
Institut d'Astrophysique de Paris (IAP)
98bis, boulevard Arago
F-75014 Paris
France